\documentclass[10pt,twocolumn]{article}

\usepackage[utf8]{inputenc}
\usepackage{titlesec}
\usepackage{booktabs}
\usepackage{graphicx}
\usepackage{amsthm}
\usepackage[colorlinks=false]{hyperref}
\usepackage[format=plain,labelfont=it]{caption}
\usepackage[left=1.5cm,right=1.5cm,top=2cm,bottom=2cm]{geometry}

\usepackage{graphicx}     
\usepackage{natbib}       
\usepackage{color}
\usepackage{amsmath} 
\usepackage{amssymb}  
\usepackage{units}
\usepackage{soul}
\usepackage{booktabs}
\usepackage{mathrsfs}
\usepackage{dsfont}
\usepackage[noend]{algpseudocode}

\titleformat{\section}{\centering\normalfont\scshape}{\arabic{section}.}{5pt}{}
\titleformat{\subsection}{\normalfont\it}{\arabic{section}.\arabic{subsection}}{5pt}{}
\titleformat{\subsubsection}{\normalfont\it}{\arabic{section}.\arabic{subsection}.\arabic{subsubsection}}{5pt}{}

\newcommand\infoFootnote[1]{%
  \begingroup
  \renewcommand\thefootnote{}\footnote{#1}%
  \addtocounter{footnote}{-1}%
  \endgroup}

\newtheorem{thm}{Theorem}
\newtheorem{cor}[thm]{Corollary}
\newtheorem{lem}[thm]{Lemma}

\newcommand{\R}{\mathbb{R}} 
 
\newcommand{\N}{\mathbb{N}}

\newcommand{\Gc}{\mathcal{G}}

\newcommand{\Xc}{\mathcal{X}}

\newcommand{\Fc}{\mathcal{F}} 
\newcommand{\Rc}{\mathcal{R}}  
\newcommand{\Tc}{\mathcal{T}}
\newcommand{\Uc}{\mathcal{U}} 
\newcommand{\Pc}{\mathcal{P}}

\newcommand{\Ac}{\mathcal{A}}
\newcommand{\Bc}{\mathcal{B}}

\newcommand{\Lc}{\mathcal{L}}

\newcommand{\eb}{\boldsymbol{e}}
\newcommand{\yb}{\boldsymbol{y}}
\newcommand{\zb}{\boldsymbol{z}}

\newcommand{\xib}{\boldsymbol{\xi}}
\newcommand{\xb}{\boldsymbol{x}}
\newcommand{\ub}{\boldsymbol{u}}
\newcommand{\bb}{\boldsymbol{b}}

\newcommand{\fb}{\boldsymbol{f}}
\newcommand{\qb}{\boldsymbol{q}}
\newcommand{\Wb}{\boldsymbol{W}}

\newcommand{\Fb}{\boldsymbol{F}}
\newcommand{\Gb}{\boldsymbol{G}}
\newcommand{\Qb}{\boldsymbol{Q}}

\newcommand{\Rb}{\boldsymbol{R}}

\newcommand{\Pb}{\boldsymbol{P}}
\newcommand{\Ab}{\boldsymbol{A}}
\newcommand{\Bb}{\boldsymbol{B}}

\newcommand{\Ib}{\boldsymbol{I}}

\newcommand{\Kb}{\boldsymbol{K}}

\newcommand{\gb}{\boldsymbol{g}}

\newcommand{\zerob}{\boldsymbol{0}}

\newcommand{\alphab}{\boldsymbol{\alpha}}
\newcommand{\betab}{\boldsymbol{\beta}}

\newcommand{\deltab}{\boldsymbol{\delta}}

\newcommand{\pib}{\boldsymbol{\pi}}

\newcommand{\Phib}{\boldsymbol{\Phi}}

\newcommand{\norm}[1]{\left\lVert#1\right\rVert}
\newcommand{\interior}{\mathrm{int}}
\newcommand{\blind}[1]{\textcolor{white}{#1}}

\DeclareMathOperator{\diag}{diag}

\title{\vspace{-2mm}\bf Error bounds for maxout neural network\\ approximations of model predictive control$^\ast$}
\author{Dieter Teichrib and Moritz Schulze Darup\vspace{2mm}}
\date{}

\begin{document}

\maketitle
              
\textbf{\textit{Abstract}.} {\bf Neural network (NN) approximations of model predictive control (MPC) are a versatile approach if the online solution of the underlying optimal control problem (OCP) is too demanding and if an exact computation of the explicit MPC law is intractable.
The drawback of such approximations is that they typically do not preserve stability and performance guarantees of the original MPC. However, such guarantees can be recovered if the maximum error with respect to the optimal control law and the Lipschitz constant of that error are known. 
We show in this work how to compute both values exactly when the control law is approximated by a maxout NN. We build upon related results for ReLU NN approximations and derive mixed-integer (MI) linear constraints that allow a computation of the output and the local gain of a maxout NN by solving an MI feasibility problem.
Furthermore, we show theoretically and experimentally that maxout NN exist for which the maximum error is zero.}
\infoFootnote{D. Teichrib and M. Schulze Darup are with the \href{https://rcs.mb.tu-dortmund.de/}{Control and~Cyber-physical Systems Group}, Faculty of Mechanical Engineering, TU Dortmund University, Germany. E-mails:  \href{mailto:dieter.teichrib@tu-dortmund.de}{\{dieter.teichrib, moritz.schulzedarup\}@tu-dortmund.de}. \vspace{0.5mm}}
\infoFootnote{\hspace{-1.5mm}$^\ast$This paper is a \textbf{preprint} of a contribution to the 22nd World Congress of the International Federation of Automatic Control 2023.}


\section{Introduction}\label{sec:Introduction}

Model predictive control (MPC) (see, e.g, \citep{Rawlings2017}) has become a standard tool for the control of dynamical systems with state and input constraints and has been successfully applied in different industrial fields (see, e.g., \citep{Qin2003} for an overview). In the classical setup, MPC requires to solve an optimization problem (OP) in every time step. For systems with a short sampling period as, e.g., in power electronics \citep{Karamanakos2020}, this can be challenging because the OP may be too complex to be solved within the sampling period.
If we consider a linear discrete-time prediction model in combination with a quadratic cost function, the resulting OP is 
a quadratic program (QP). In principle, we can compute the solution of the parametric QP offline for all feasible states. This results in an explicit control law with a piecewise affine (PWA) input-output relation defined on a polyhedral partition of the state space \citep{Bemporad2002}. Given the explicit control law, the online computational effort reduces to the evaluation of the PWA function. However, the number of polyhedral regions may grow exponentially with the state dimension and the number of constraints in the OP. Hence, exactly computing the explicit MPC law becomes untractable for complex systems. 
As a consequence, various techniques have been developed to approximate the control law in MPC \citep{Jones2009,Bemporad2003}. In this context, neural networks (NN) are very popular \citep{Chen2018,Karg2020,Chen2022} since they can approximate a large class of functions, including PWA functions, with arbitrary accuracy \citep{Hornik1989}. In addition, besides their computational demanding offline training, NN are typically fast to evaluate online, which is essential if they are used as controllers. Moreover, some types of NN share the PWA structure of the control law \citep{SchulzeDarup_2020_ECC_ANN,Hanin2017,Arora2016}, making them the perfect choice for approximating MPC. Unfortunately, in general, stability cannot be guaranteed for the approximated controllers. One possibility to recover stability and recursive feasibility is to project the output of the NN onto a suitable set as, e.g., in \citep{Paulson2020,Chen2018}. Alternatively, the output of the NN can be used as an initial guess for a solver and not directly for control \citep{Chen2022}. The drawback of both approaches is that they require an additional optimization-based computation step online. Ideally, the NN can be used as a controller without additional online computation, while still providing stability guarantees. In \citep{Fabiani2021}, it is proven that this is possible if the maximum error with respect to the optimal control law and the Lipschitz constant of the corresponding error function are known. 
Both values can indeed be computed exactly presupposed the output and the local gain of the used NN can be computed by solving a mixed-integer (MI) feasibility problem. This is known to be possible for NN using rectified linear units (ReLU) as activation functions \citep[Thm.~6.1]{Fabiani2021}. 

In the work at hand, we will extend the result of \citep{Fabiani2021} by showing that the maximum error and the Lipschitz constant of the error can also be computed exactly for a controller approximation based on a maxout NN. Since maxout NN include other NN with PWA input-output relation such as, e.g., ReLU and leaky ReLU, as a special case, the results provide a generalization to a broader class of PWA NN. Furthermore, we use the PWA structure of maxout NN to compute NN that exactly describe MPC control laws and validate experimentally that these exact maxout NN indeed lead to a maximum error and Lipschitz constant of zero.      

The paper is organized as follows. In the remainder of this section, we introduce relevant notation. In Section~\ref{sec:fundamentals}, we summarize some basics on MPC as well as PWA NN and describe concepts for approximating MPC in more depth. Section~\ref{sec:MaxoutNNaMPC} is devoted to our main result, i.e., the computation of the maximum error and the Lipschitz constant of the error function related to maxout NN approximating MPC. The obtained method is applied to various maxout NN approximations of MPC laws in Section~\ref{sec:examples}. Finally, conclusions and an outlook are given in Section~\ref{sec:conclusions}.   

\subsection{Notation}

We will denote the index set containing $p_i\in \N$ integers starting at $p_i(l-1)+1, \ l \in \N$ by 
\begin{equation*}
    \Ac^{(i)}_l:=\{p_i(l-1)+1,\dots,p_i l\}.
\end{equation*}
For vectors $\xb\in\R^n$ we denote the $i$-th element by $\xb_i$ and the elements between the indices $a_1$ and $a_2>a_1$ by $\xb_{a_1:a_2}$. For matrices $\Kb\in\R^{m \times n}$ we denote the element in the $i$-th row and $j$-th column by $\Kb_{i,j}$, the $i$-row and $j$-th column by $\Kb_{i,:}$ and $\Kb_{:,j}$, respectively. If we only write $\Kb_i$ then we refer to the $i$-th row. A block diagonal matrix is defined as 
\begin{equation*}
    \diag(\alphab_1,\dots,\alphab_N):=
    \begin{pmatrix}
        \alphab_1 & \zerob & \hdots & \zerob \\
        \zerob & \ddots & & \vdots \\
        \vdots &  &  & \zerob \\
        \zerob & \hdots & \zerob & \alphab_{w_i}\\
    \end{pmatrix},
\end{equation*}
with $\alphab \in \R^{w_i \times p_i}$. A continuous function $\Fb(\xb):\Pc \subset \R^n \rightarrow\R^m$ of the form
\begin{equation}\label{eq:PWA_f}
\Fb(\xb) = \left\{
\begin{array}{cc}
 \Gb^{(1)} \xb + \gb^{(1)} & \text{if}\,\,\,\xb\in\Pc^{(1)}, \\
\vdots & \vdots \\
 \Gb^{(s)} \xb+ \gb^{(s)} & \text{if}\,\,\,\xb\in\Pc^{(r)}, \\
\end{array}
\right. 
\end{equation} 
with a polyhedral partition $\Pc=\cup_{i=1}^r \Pc^{(i)}$ and $\interior(\Pc^{(i)})\cap\interior(\Pc^{(j)})=\emptyset \ \forall i\neq j$ is denoted as piecewise affine (PWA) function. We further define the local gain $\Kb(\xb):\cup_{i=1}^r \interior(\Pc^{(i)})\rightarrow\R^m$ of a PWA function as
\begin{equation}\label{eq:K_NN_delta1}
    \Kb(\xb) := \left\{
\begin{array}{cc}
 \Gb^{(1)} & \text{if}\,\,\,\xb\in\interior(\Pc^{(1)}), \\
\vdots & \vdots \\
 \Gb^{(s)} & \text{if}\,\,\,\xb\in\interior(\Pc^{(r)}). \\
\end{array}
\right.
\end{equation}

\section{Fundamentals of MPC and NN}\label{sec:fundamentals}

\subsection{Model predicitive control}

Model predictive control (MPC) for linear discrete-time systems builds on solving an optimal control problem (OCP) of the form
\begin{align}
\label{eq:OCP}
V_N(\xb) := \!\!\!\!\min_{\substack{\hat{\xb}(0),...,\hat{\xb}(N)\\ \hat{\ub}(0),...,\hat{\ub}(N-1)}} 
\!\!\!\!\!\!\!\!\!\!\!\!\!\!\!\!\!\!\!\! & 
\,\,\,\,\,\,\,\,\,\,\,\,\,\,\,\,\,\,\,\,\,
\varphi( \hat{\xb}(N)) + \! \sum_{\kappa=0}^{N-1} \ell(\hat{\xb}(\kappa),\hat{\ub}(\kappa))   \\
\nonumber
\text{s.t.} \quad \quad  \hat{\xb}(0)&=\xb, \\
\nonumber
 \hat{\xb}(\kappa+1)&=\Ab\,\hat{\xb}(\kappa) + \Bb \hat{\ub}(\kappa), \quad\!\!\forall \kappa \in \{0,...,N-1\}, \\
 \nonumber
\left(\hat{\xb}(\kappa),\hat{\ub}(\kappa)\right) & \in \Xc \times \Uc, \quad\hspace{14.9mm}\forall \kappa \in \{0,...,N-1\}, \\
 \nonumber
 \hat{\xb}(N) & \in \Tc
 \nonumber
\end{align}
in every time step $k\in\N$ for the current state ${\xb=\xb(k)}$. 
Here, $N\in\N$ refers to the prediction horizon and 
\begin{equation}
\varphi(\xb):= \xb^\top \Pb \xb \quad \text{and} \quad \ell(\xb,\ub):=\xb^\top \Qb \xb + \ub^\top \Rb \ub
\end{equation}
denote the terminal and stage cost, respectively,  where the weighting matrices  $\Pb$, $\Qb$, and $\Rb$ are positive (semi\nobreakdash-) definite. The dynamics of the linear prediction model are described by $\Ab\in \R^{n \times n}$ and ${\Bb \in \R^{n\times m}}$. State and input constraints can be incorporated via the polyhedral sets $\Xc$ and $\Uc$. Finally, the terminal set $\Tc$ allows to enforce closed-loop stability 
(see \citep{Mayne2000} for details).
The resulting control law $\pib:\Fc_N \rightarrow \Uc$ is defined~as
\begin{equation}
\label{eq:gMPC}
\pib(\xb):=\hat{\ub}^\ast(0),
\end{equation}
where $\Fc_N$ denotes the feasible set of~\eqref{eq:OCP} and where $\hat{\ub}^\ast(0)$ refers to the first element of the optimal input sequence.
For the considered setup it is well known that $\pib(\xb)$ is a PWA function \citep[Thm.~4]{Bemporad2002} of the form \eqref{eq:PWA_f} with $\Gb^{(i)}=\Kb^{(i)}$, $\gb^{(i)}=\bb^{(i)}$, $r=r_{\text{MPC}}$, polyhedral sets $\Pc^{(i)}=\Rc^{(i)} \ \forall i \in \{1,\dots, r_{\text{MPC}}\}$ and local gain $\Kb_{\text{MPC}}(\xb)$. 

\subsection{Neural networks with piecewise affine activations}

In general, a feed-forward-NN with $\ell \in\N$ hidden layers and $w_i$ neurons in layer $i$ can be written as a composition of the form
\begin{equation}\label{eq:NN}
    \Phib(\xb)=\fb^{(\ell+1)}\circ \gb^{(\ell)}\circ \fb^{(\ell)}\circ \dots \circ \gb^{(1)}\circ \fb^{(1)}(\xb).
\end{equation}
Here, the functions $\fb^{(i)}: \R^{w_{i-1}} \rightarrow \R^{p_i w_i}$ for $i \in \{1,\dots,\ell\}$ refer to preactivations, where the parameter $p_i\in \N$ allows to consider ``multi-channel'' preactivations as required for maxout (see \citep{Goodfellow2013}). Moreover, $\gb^{(i)}: \R^{p_i w_i} \rightarrow \R^{w_i}$ stand for activation functions and $\fb^{(\ell+1)}: \R^{w_{\ell}} \rightarrow \R^{w_{\ell+1}}$ reflects  postactivation.
The functions $\fb^{(i)}$ are typically affine, i.e.,
\begin{equation}\label{eq:NN_fi}
    \fb^{(i)}(\yb^{(i-1)})=\Wb^{(i)}\yb^{(i-1)}+\bb^{(i)},
\end{equation}
where $\Wb^{(i)}\in\R^{p_i w_i\times w_{i-1}}$ is a weighting matrix, $\bb^{(i)} \in \R^{p_i w_i}$ is a bias vector, and 
 $\yb^{(i-1)}$ denotes the output of the previous layer with $\yb^{(0)}:=\xb \in \R^n$. 

Now, various activation functions have been proposed. As already stated in the introduction, we here focus on PWA activation functions, i.e., we consider the ReLU activation function
\begin{equation}
    \label{eq:ReLU}
    \gb_{\text{ReLU}}^{(i)}(\zb^{(i)})=\max\left\{\zerob,\zb^{(i)}\right\}:=\begin{pmatrix}
        \max\big\{0,\zb_1^{(i)}\big\} \\
        \vdots \\
        \max\big\{0,\zb_{w_i}^{(i)}\big\}
    \end{pmatrix}
\end{equation}
and the maxout activation function
\begin{equation}
    \label{eq:maxout}
    \gb_{\text{max}}^{(i)}(\zb^{(i)})=
    \begin{pmatrix}
        \max \limits_{1\leq j \leq p_i}\big\{\zb_j^{(i)}\big\} \\
        \vdots \\
        \max \limits_{p_i(w_i-1)+1 \leq j \leq p_i w_i}\big\{\zb_{j}^{(i)}\big\}
    \end{pmatrix},
\end{equation}
where we use the shorthand notation
$$
\max\limits_{1\leq j \leq p_i}\big\{\zb_j^{(i)}\big\}:=\max\big\{\zb_1^{(i)},\dots,\zb_{p_i}^{(i)}\big\}.
$$
We will refer to the resulting NN as ReLU NN and maxout NN, respectively. The proof of \citep[Thm.~4.3]{Goodfellow2013} shows that maxout NN are PWA functions of the form \eqref{eq:PWA_f} with $\Gb^{(i)}=\Kb_{\text{NN}}^{(i)}$, $\gb^{(i)}=\bb_{\text{NN}}^{(i)}$, $r=r_{\text{NN}}$, polyhedral sets $\Pc^{(i)}=\Rc_{\text{NN}}^{(i)} \ \forall i \in \{1,\dots, r_{\text{NN}}\}$ and local gain $\Kb_{\text{NN}}(\xb)$. The number of parameters needed to describe a NN is
\begin{equation*}
    \#_p:=\sum\limits_{i=1}^{\ell}(w_{i-1}+1)p_i w_i + (w_{\ell}+1) w_{\ell+1},
\end{equation*}
where $\ell$, $p_i$ with $i \in \{1,\dots,\ell\}$ and $w_i$ with $i \in \{1,\dots,\ell+1\}$ describe the topology of the NN.

\subsection{Approximate MPC}\label{subsec:aproxMPC}

The use of approximate MPC is particularly useful, if the OCP \eqref{eq:OCP} is too complex to be solved online and the explicit solution has too many regions and thus a large memory footprint (\citep{Kvasnica2012}). In this case, the exact control law may be approximated by a function that is fast to evaluate and has a small memory footprint. A promising candidate for such a function is an NN, as it combines both required properties and is a universal function approximator \citep[Thm.~2.4]{Hornik1989}. In addition, due to the common PWA structure of some NN \citep[Thm.~2]{Hanin2017}, \citep[Thm.~2.1]{Arora2016} and the control law \citep[Thm.~4]{Bemporad2002}, they seem to be a natural choice. In fact, for a suitable choice of the weighting matrices and bias vectors, ReLU NN can represent the control law exactly \citep[Thm.~1]{SchulzeDarup_2020_ECC_ANN}, \citep[Thm.~1]{Karg2020}. Unfortunately, despite the ability of NN to exactly represent the control law, the approximated version of the control law typically does not preserve desirable properties of the MPC, such as stability and performance. Crucial for preserving these properties is the error function
\begin{equation}\label{eq:error}
    \eb(\xb):=\pib(\xb)-\Phib(\xb).
\end{equation}
More precisely we have to compute the maximum error
\begin{equation}\label{eq:MaxError}
    \overline{e}_\alpha:=\max\limits_{x \in \Xc} \norm{\eb(\xb)}_\alpha
\end{equation}
and the $\alpha$-Lipschitz constant of the error
\begin{equation*}
    \Lc_\alpha(\eb,\Xc):=\sup\limits_{\xb\neq \yb \in \Xc} \frac{\norm{\eb(\xb)-\eb(\yb)}_\alpha}{\norm{\xb-\yb}_\alpha}.
\end{equation*}
According to \citep[Prop.~3.4]{Gorokhovik1994}, the $\alpha$-Lipschitz constant of a PWA function is equal to the maximum $\alpha$-norm of the local gain. If we consider a PWA NN, the error is as difference of two PWA functions also PWA \citep[Prop.~1.1]{Gorokhovik1994} and the $\alpha$-Lipschitz constant is thus 
\begin{equation}\label{eq:Lipschitz_e}
    \Lc_\alpha(\eb,\Xc)=\max\limits_{\xb \in \Xc}\norm{\Kb_{\text{MPC}}(\xb)-\Kb_{\text{NN}}(\xb)}_\alpha.
\end{equation}
Now, if $e_\alpha$ and $\Lc_\alpha(\eb,\Tc)$ are below certain values, specified in \citep[Eq.~(24)--(25)]{Fabiani2021}, then the closed-loop system with the approximated NN controller converges exponentially to the origin according to \citep[Thm.~3.4]{Fabiani2021}. The problem at this point is that during the training of a NN, the error \eqref{eq:error} is only evaluated at discrete samples $(\xb_i^\top \ \pib(\xb_i)^\top)$ of the control law, and the parameters of the NN are chosen such that the mean squared error (MSE)
\begin{equation}
    \label{eq:MSE}
    \hat{e}^2:=\frac{1}{D}\sum\limits_{i=1}^D \norm{\pib(\xb_i)-\Phib(\xb_i)}_2^2
\end{equation}
over the $D \in \N$ training samples is minimized. We thus do not have any guarantees for the error at points $\xb\in\Fc_N$ not included in the training samples. Moreover, a low MSE does not necessarily mean that the values of \eqref{eq:MaxError} and \eqref{eq:Lipschitz_e} are low. Therefore, to certify stability of the closed-loop system with a pre-trained NN we need a way to compute these values exactly. For $\alpha=\{1,\infty\}$ this is possible by solving a mixed-integer linear program (MILP) \citep[Thm.~6.1]{Fabiani2021} if both the output $\Phib(\xb)$ and the local gain $\Kb_{\text{NN}}(\xb)$ of the NN can be computed by solving an MI feasibility problem. Which is proven for ReLU NN in \citep[Thm.~6.1]{Fabiani2021}. In the remainder of this paper, we will extend the results to maxout NN.      

\section{Maxout neural networks for approximate MPC}\label{sec:MaxoutNNaMPC}
 
In most of the recent work where the control law \eqref{eq:gMPC} is approximated by an NN, ReLU NN are used as function approximators (see, e.g., \citep{Chen2022,Drummond2022,Karg2020}). Maxout NN are rarely considered in this context, although they offer a number of advantages. First, the fact that ReLU NN can represent every PWA function exactly is often used as justification for their use to approximate the PWA control law. However, most ReLU NN that allow an exact description are based on the representation of PWA functions as a sum of the type
\begin{equation}\label{eq:FHat}
    \hat{F}(\xb) =  \sum\limits_{i=1}^M \sigma_i \max\limits_{1\leq j \leq J}\{ \betab^{(i)}_j \xb + \gamma^{(i)}_j \}.
\end{equation}
From \citep{Wang2005} and \citep{Kripfganz1987} it is known for which $M$ and $J$ we can find parameters $\sigma_i$, $\betab_j$ and $\gamma_j$ such that $\hat{F}(\xb)=F(\xb)$ holds for $m=1$. Since \eqref{eq:FHat} is a maxout NN with $\ell=1$, $w_1=M$, $p_1=J$, $w_2=1$, these results can be used directly to find suitable topologies for maxout NN that allow an exact description of the control law. For ReLU NN these results are not directly applicable. Therefore, \eqref{eq:FHat} is decomposed in, e.g., \citep{Hanin2017,Arora2016} to find a ReLU topology that allows an exact description. Such a topology is used in, e.g., \citep[Thm.~1]{Karg2020} to represent the control law. The decomposition step typically leads to a more conservative ReLU topology in terms of number of layers $\ell$ and neurons per layer $w_i$ compared to a maxout NN that directly represents \eqref{eq:FHat}. Another advantage of the maxout activation is that it trivially include the ReLU activation as a special case. In fact, a ReLU activation is a maxout activation with $p_i=2$ where every second affine segment is set to zero (cf. \eqref{eq:ReLU} and \eqref{eq:maxout}). Thus an approach that allow the computation of \eqref{eq:MaxError} and \eqref{eq:Lipschitz_e} for maxout NN is also applicable to ReLU NN and hence extends the known results.

\subsection{Exact maxout neural networks}\label{subsec:exactNN}

\begin{cor}\label{cor:exactMaxout}
Let $F(\xb)$ be an arbitrary PWA function of the form \eqref{eq:PWA_f} with one dimensional output, i.e, $m=1$. Then for a maxout NN $\Phi(\xb)$ with $\ell=1$, $w_2=1$ and $\bb^{(2)}=0$ there exist parameters 
\begin{itemize}
    \item[(i)] $p_1\!\in\!\N$, $\Wb^{(1)}\!\in\!\R^{2 p_1 \times n}$, $\bb^{(1)}\!\in\!\R^{2 p_1 \times 1}$ and $\Wb^{(2)}\!\in\!\R^{1 \times 2}$
\end{itemize}
and
\begin{itemize}
    \item[(ii)] $w_1\!\in\!\N$, $\Wb^{(1)}\!\in\!\R^{w_1 (n+1) \times n}$, $\bb^{(1)}\!\in\!\R^{w_1 (n+1) \times 1}$ and $\Wb^{(2)}\!\in\!\R^{1 \times w_1}$
\end{itemize}
with $\Phi(\xb)=F(\xb)$.
\end{cor}
\begin{proof}
Since the resulting maxout NN are of the form \eqref{eq:FHat} with $J=p_1$, $M=2$ for (i) and $J=n+1$, $M=w_1$ for (ii), the proof follows from \citep[Lem.~1]{Kripfganz1987} and \citep[Thm.~1]{Wang2005}, respectively.
\end{proof}

Corollary~\ref{cor:exactMaxout} provides two different topologies (i) and (ii) for maxout NN with one hidden layer that can represent every PWA function with one dimensional output exactly, if $p_1$ and $w_1$, respectively are chosen large enough. Although the results are formulated for $m=1$ they can also be applied to PWA functions with $m>1$ by applying Corollary~\ref{cor:exactMaxout} to every dimension of the output individually. Thus maxout NN can represent the PWA control law \eqref{eq:gMPC} for arbitrary state and input dimension.   

\subsection{Maxout neural networks as MILP}
 
In this section, we will derive our main result, which is the computation of the output and the local gain of NN with maxout activation by solving an MI feasibility problem. This result then allows to compute \eqref{eq:MaxError} and \eqref{eq:Lipschitz_e} for the case where the control law \eqref{eq:gMPC} is approximated by a maxout NN. Both values are according to the descriptions in Section~\ref{subsec:aproxMPC} sufficient to prove stability of the closed-loop system. Moreover, these values provide a more profound way to evaluate the success of the training than by just considering the error at the training samples \eqref{eq:MSE} as in, e.g., \citep{Karg2020,Teichrib2021}. We will derive our results based on the observation that the output of a maxout NN with $\ell$ hidden layers can be modeled by the recursion
\begin{align}
    \nonumber
    \yb^{(0)}&=\xb,\\
    \nonumber
    \yb^{(i)}&=\Delta^{(i)}(\Wb^{(i)}\yb^{(i-1)}+\bb^{(i)}), \ 1\leq i \leq \ell,\\
    \label{eq:MaxoutRec}
    \Phib(\xb)&=\Wb^{(\ell+1)}\yb^{(\ell)}+\bb^{(\ell+1)}
\end{align}
where $\Delta^{(i)}$ is a block diagonal matrix of the form 
\begin{equation}\label{eq:Delta^i}
    \Delta^{(i)}:=\diag(\deltab^{(i)}_{1:p_i},\dots,\deltab^{(i)}_{p_i(w_i-1)+1:p_i w_i})
\end{equation}
with binary variables $\deltab^{(i)} \in \mathbb{B}^{1 \times p_i w_i}$. The matrix $\Delta^{(i)}$ is such that for all elements $\deltab^{(i)}_j$ the logical implication
\begin{align}
    \nonumber
    &[\deltab^{(i)}_{k_s}=1] \Longleftrightarrow [\Wb^{(i)}_j \yb^{(i-1)} + \bb^{(i)}_j \leq \Wb^{(i)}_{k_s} \yb^{(i-1)} + \bb^{(i)}_{k_s},\\
    \nonumber
    &\hspace{24mm} \forall j \in \Ac^{(i)}_s \setminus k_s, \ k_s \in \Ac^{(i)}_s ],\\
    \label{eq:ImpMaxoutNN}
    &\forall s \in \{1,\dots,w_i\}, \ \forall i \in \{1,\dots,\ell\}.
\end{align}
holds. Thus, for every neuron in layer $i$ the matrix $\Delta^{(i)}$ selects the largest affine segment among all $p_i$ segments. With the results from \citep[Sec.~2]{Fischetti2018} we can model the logical implication \eqref{eq:ImpMaxoutNN} by the following MI linear constraints
\begin{align}
    \nonumber
    \qb^{(i)}_s &\leq \Wb^{(i)}_j \qb^{(i-1)} + \bb^{(i)}_j + \overline{b}^{(i)} (1-\deltab^{(i)}_j), \\
    \nonumber
    -\qb^{(i)}_s &\leq -\Wb^{(i)}_j \qb^{(i-1)} - \bb^{(i)}_j - \varepsilon(1-\deltab^{(i)}_j), \\
    \nonumber
    \qb^{(0)}&= \xb, \\
    \nonumber
    \sum\limits_{\jmath\in\Ac^{(i)}_s} &\deltab^{(i)}_\jmath =1, \quad  \\
    \label{eq:MI_Constraints}
    \forall j &\in \Ac^{(i)}_s, \ \forall s \in \{1,\dots,w_i\}, \ \forall i \in \{1,\dots,\ell\},
\end{align}
with a constant upper bound $\overline{b}^{(i)}\in\R$ and a small $\varepsilon \geq 0$. The variables $\qb^{(i)} \in \R^{w_i}$ and $\deltab^{(i)}$ are real and binary optimization variables, respectively. 
\begin{lem}\label{lem:MI_Maxout}
Let $\qb^{(i)}$ and $\deltab^{(i)}$ be such that the constraints \eqref{eq:MI_Constraints} with $\varepsilon=0$ hold. Then, the output of the maxout NN \eqref{eq:NN} is given by
\begin{equation}\label{eq:Maxout_q}
    \Phib(\xb)=\Wb^{(\ell+1)}\qb^{(\ell)}+\bb^{(\ell+1)}.
\end{equation}
\end{lem}
\begin{proof}
If we can show that 
\begin{equation}\label{eq:q=y}
    \qb^{(i)}=\yb^{(i)}
\end{equation}
holds for all $i\in\{0,\dots,\ell\}$ then \eqref{eq:Maxout_q} holds according to \eqref{eq:NN} and \eqref{eq:NN_fi}. We prove this by induction. The base case $i=0$ is true by assumption since we have $\qb^{(0)}=\xb=\yb^{(0)}$. Moreover, the constraints \eqref{eq:MI_Constraints} are such that for every index set $\Ac^{(i)}_s$ there exists exactly one $k_s \in \Ac^{(i)}_s$ with $\deltab^{(i)}_{k_s}=1$ and $\deltab^{(i)}_j=0, \forall j \in \Ac^{(i)}_s \setminus k_s$. Thus if we assume that the induction hypothesis \eqref{eq:q=y} is true for one $i=t$ we obtain for $i=t+1$ the constraints
\begin{align*}
    \qb^{(t+1)}_s &= \Wb^{(t+1)}_{k_s} \yb^{(t)} + \bb^{(t+1)}_{k_s}, \ {k_s} \in \Ac^{(t+1)}_s,\\
    \qb^{(t+1)}_s &\leq \Wb^{(t+1)}_j \yb^{(t)} + \bb^{(t+1)}_j + \overline{b}^{(t+1)}, \ \forall j \in \Ac^{(t+1)}_s \setminus k_s, \\
    \qb^{(t+1)}_s &\geq \Wb^{(t+1)}_j \yb^{(t)} + \bb^{(t+1)}_j, \ \forall j \in \Ac^{(t+1)}_s \setminus k_s,\\
    \forall s &\in \{1,\dots,w_{t+1}\}.
\end{align*}
The first two constraints always hold for a large $\overline{b}^{(i)}$. Hence the constraints \eqref{eq:MI_Constraints} imply
\begin{align*}
    &\qb^{(t+1)}_s=\Wb^{(t+1)}_{k_s} \yb^{(t)} + \bb^{(t+1)}_{k_s} \geq \Wb^{(t+1)}_j \yb^{(t)} + \bb^{(t+1)}_j, \\ &\forall j \in \Ac^{(t+1)}_s \setminus k_s, \ \forall s \in \{1,\dots,w_{t+1}\}.
\end{align*}
Which is exactly the relation \eqref{eq:ImpMaxoutNN} for $i=t+1$, i.e., the largest affine segment of the $s$-th neuron is the $k_s$-th segment. Thus the relation
\begin{align*}
    \begin{pmatrix}\!
        \max \limits_{j\in \Ac^{(t+1)}_1}\!\!\big\{\Wb^{(t+1)}_j \yb^{(t)}\!+\!\bb^{(t+1)}_j\big\} \\
        \vdots \\
        \max \limits_{j\in \Ac^{(t+1)}_{w_{t+1}}}\!\!\big\{\Wb^{(t+1)}_j \yb^{(t)}\!+\!\bb^{(t+1)}_j\big\}
    \!\end{pmatrix}\!\!&=\!\!
    \begin{pmatrix}
        \Wb^{(t+1)}_{k_1} \yb^{(t)}\!+\!\bb^{(t+1)}_{k_1} \\
        \vdots \\
        \Wb^{(t+1)}_{k_{w_{t+1}}} \yb^{(t)}\!+\!\bb^{(t+1)}_{k_{w_{t+1}}}
    \end{pmatrix}\\
    \Longleftrightarrow \yb^{(t+1)}\!&=\!\qb^{(t+1)}
\end{align*}
hold. This proves that \eqref{eq:q=y} holds for all $i\in\{0,\dots,\ell\}$. With 
\begin{equation*}
    \Phib(\xb)=\fb^{(\ell+1)}(\yb^{(\ell)})=\fb^{(\ell+1)}(\qb^{(\ell)})
\end{equation*}
the output of the maxout NN can be computed via \eqref{eq:Maxout_q}.
\end{proof}
Next, we will derive MI linear constraints for the computation of the local gain of a maxout NN. Applying the chain rule for derivative to the recursive formula \eqref{eq:MaxoutRec} leads to $\Kb_{\text{NN}}:\Gc \rightarrow \R^m$
\begin{equation}\label{eq:K_NN_delta}
    \Kb_{\text{NN}}(\xb) := \nabla \Phib(\xb)^\top =\Wb^{(\ell+1)} \prod\limits_{i=1}^{\ell} \Delta^{(i)}\Wb^{(i)}
\end{equation}
as expression for the local gain. The gradient is well-defined everywhere except on the boundaries between two neighboring affine segments of the PWA function represented by the maxout NN. These boundaries are given by
\begin{align}
    \nonumber
    \Bc :=& \bigcup\limits_{i=1}^\ell \bigcup\limits_{l=1}^{w_i} \Bc^{(i)}_l \ \text{with}\\
    \nonumber
    \Bc^{(i)}_l:=&\{ \xb \in \R^n \ | \ \exists k \in \Ac^{(i)}_l \exists \tilde{k} \in \Ac^{(i)}_l \setminus k, \tilde{\deltab}^{(i)}_k=1,\\
    & \ \Wb^{(i)}_k \yb^{(i-1)} + \bb^{(i)}_k=\Wb^{(i)}_{\tilde{k}} \yb^{(i-1)} + \bb^{(i)}_{\tilde{k}}\},
\end{align}
where $\tilde{\deltab}^{(i)}_k$ is subject to the constraints \eqref{eq:MI_Constraints} with $\deltab^{(i)}=\tilde{\deltab}^{(i)}$ and $\varepsilon=0$. We denote the set where the gradient is well-defined by
\begin{equation*}
    \Gc:=\R^n \setminus \Bc.
\end{equation*}
Since the matrix $\Delta^{(i)}$ selects the active, i.e., the largest, affine segment of the maxout neurons, we can model the computation of the local gain $\Kb_{\text{NN}}(\xb)$ by the additional MI linear constraints 
\begin{align}
    \nonumber
    &\underline{w}^{(i)}_{h,r} \deltab^{(i)}_h \leq \tilde{\xib}^{(i)}_{h,r} \leq \overline{w}^{(i)}_{h,r} \deltab^{(i)}_h, \\
    \nonumber
    &-\overline{w}^{(i)}_{h,r} (1-\deltab^{(i)}_h) \leq \tilde{\xib}^{(i)}_{h,r} - \tilde{\Wb}^{(i)}_{h,r} \leq -\underline{w}^{(i)}_{h,r} (1-\deltab^{(i)}_h),\\
    \nonumber
    &\xib^{(i)}_{1,r}=\sum\limits_{\tilde{h}=1}^{p_i}\tilde{\xib}^{(i)}_{\tilde{h},r}, \ \dots, \ \xib^{(i)}_{w_i,r}=\sum\limits_{\tilde{h}=p_i(w_i-1)+1}^{p_i w_i}\tilde{\xib}^{(i)}_{\tilde{h},r}, \\
    \nonumber
    &\tilde{\Wb}^{(i)}=\Wb^{(i)} \xib^{(i-1)}, \quad \xib^{(0)}=\Ib\\
    \label{eq:MI_Constraints_Gain}
    &\forall h \in \{1,\dots,p_i w_i\},  \forall r \in \{1,\dots,n\},  \forall i \in \{1,\dots,\ell\},
\end{align}
with constant lower and upper bounds $\underline{w}^{(i)}_{h,r}$ and $\overline{w}^{(i)}_{h,r}$, respectively. The MI constraints \eqref{eq:MI_Constraints_Gain} are similar to those used in \citep[Eq.~20]{Fabiani2021}, except that here we need additional auxiliary variables to account for the $p_i$ different affine segments in each maxout neuron.  

\begin{lem}\label{lem:MI_Maxout_gain}
Let $\qb^{(i)}$, $\deltab^{(i)}$, $\tilde{\Wb}^{(i)}$ and $\tilde{\xib}^{(i)}$ be such that the constraints \eqref{eq:MI_Constraints} and \eqref{eq:MI_Constraints_Gain} with $\varepsilon>0$ hold. Then, the local gain of a maxout NN is given by
\begin{equation}\label{eq:K_NN_xi}
    \Kb_{\text{NN}}(\xb)=\Wb^{(\ell+1)} \xib^{(\ell)}.
\end{equation}
\end{lem}
\begin{proof}
We first proof that \eqref{eq:K_NN_xi} is well-defined for $\xb \in \Gc$. Since we know that \eqref{eq:q=y} holds for all $i\in\{0,\dots,\ell\}$, we can use the same argumentation as in the proof of Lemma~\ref{lem:MI_Maxout} to rewrite the constraints as follows 
\begin{align}
    \nonumber
    \yb^{(i)}_s &= \Wb^{(i)}_{k_s} \yb^{(i-1)} + \bb^{(i)}_{k_s}, \ k_s \in \Ac^{(i)}_s,\\
    \nonumber
    \yb^{(i)}_s &\leq \Wb^{(i)}_j \yb^{(i-1)} + \bb^{(i)}_j + \overline{b}^{(i)}, \ \forall j \in \Ac^{(i)}_s \setminus k_s, \\
    \nonumber
    \yb^{(i)}_s &\geq \Wb^{(i)}_j \yb^{(i-1)} + \bb^{(i)}_j + \varepsilon, \ \forall j \in \Ac^{(i)}_s \setminus k_s,\\
    \label{eq:MI_Constraints_proof}
    \forall s &\in \{1,\dots,w_{i}\}, \ \forall i \in \{1,\dots,\ell\}.
\end{align}
Again, the first two constraints always hold, if $\overline{b}^{(i)}$ is chosen large enough. Thus the feasibility only depends on the last constraint
\begin{align*}
    &\Wb^{(i)}_{k_s} \yb^{(i-1)}\!+\!\bb^{(i)}_{k_s}\!-\!\Wb^{(i)}_j \yb^{(i-1)}\!-\!\bb^{(i)}_j \geq \varepsilon,\ \forall j \in \Ac^{(i)}_s \setminus k_s,
\end{align*}
with $s$ and $i$ as in \eqref{eq:MI_Constraints_proof}. Moreover, we can assume without loss of generality that the left-hand side is positive, because if this is not the case for some $j$, the role of that $j$ can be changed with the role of $k_s$. 
Hence the constraints are infeasible if and only if
\begin{equation}\label{eq:kMinusjLessEpsilon}
    \exists j\!\in\!\Ac^{(i)}_s \setminus k_s\!:\! \left| (\Wb^{(i)}_{k_s} - \Wb^{(i)}_j) \yb^{(i-1)} + \bb^{(i)}_{k_s} - \bb^{(i)}_j \right|\!<\!\varepsilon
\end{equation}
holds. If we define the set
\begin{align*}
    \overline{\Bc}:=& \bigcup\limits_{i=1}^\ell \bigcup\limits_{l=1}^{w_i} \overline{\Bc}^{(i)}_l \ \text{with}\\
    \overline{\Bc}^{(i)}_l:=& \{\xb \in \R^n \ | \ \exists k \in \Ac^{(i)}_l \exists \tilde{k} \in \Ac^{(i)}_l \setminus k, \tilde{\deltab}^{(i)}_k=1,\\
    & \ | (\Wb^{(i)}_k - \Wb^{(i)}_{\tilde{k}}) \yb^{(i-1)} + \bb^{(i)}_k - \bb^{(i)}_{\tilde{k}} | < \varepsilon \},
\end{align*}
we have $\xb \in \Bc \subset \overline{\Bc} \Rightarrow \xb \in \overline{\Bc} \Leftrightarrow \eqref{eq:kMinusjLessEpsilon}$, i.e., the constraints are feasible for $\xb \in \R^n\setminus\overline{\Bc}$. For small $\varepsilon$ we have $\overline{\Bc} \approx \Bc$. Thus the domain of \eqref{eq:K_NN_delta} is approximately the feasible set of \eqref{eq:MI_Constraints}. In addition, the solution to \eqref{eq:MI_Constraints} is unique. We prove this by contradiction and assume that there exist multiple solutions and thus there exists a $\tilde{k}_s$ with 
\begin{equation*}
    \deltab^{(i)}_{\tilde{k}_s}=1, \quad \tilde{k}_s \in \Ac^{(i)}_s \setminus k_s.
\end{equation*}
This would imply 
\begin{equation*}
    \Wb^{(i)}_{\tilde{k}_s} \yb^{(i-1)} + \bb^{(i)}_{\tilde{k}_s} \geq \Wb^{(i)}_{k_s} \yb^{(i-1)} + \bb^{(i)}_{k_s} + \varepsilon.
\end{equation*}
Further, since for feasible $\xb$ the inequality
\begin{equation*}
    \Wb^{(i)}_{k_s} \yb^{(i-1)} + \bb^{(i)}_{k_s} - \varepsilon \geq \Wb^{(i)}_{\tilde{k}_s} \yb^{(i-1)} + \bb^{(i)}_{\tilde{k}_s}
\end{equation*}
holds, we obtain the contradiction
\begin{align*}
    \Wb^{(i)}_{k_s} \yb^{(i-1)} + \bb^{(i)}_{k_s} + \varepsilon &\leq \Wb^{(i)}_{\tilde{k}_s} \yb^{(i-1)} + \bb^{(i)}_{\tilde{k}_s} \\
    &\leq \Wb^{(i)}_{k_s} \yb^{(i-1)} + \bb^{(i)}_{k_s} - \varepsilon.
\end{align*}
Thus the solution exists and is unique for $\xb \in \R^n \setminus \overline{\Bc}$. Next, we prove that \eqref{eq:K_NN_xi} holds for feasible $\xb$, starting with the case $\deltab^{(i)}_h=0$, where we have
\begin{align*}
    &\tilde{\xib}^{(i)}_{h,r}=0 \ \text{and} \ \underline{w}^{(i)}_{h,r}\leq \tilde{\Wb}^{(i)}_{h,r} \leq \overline{w}^{(i)}_{h,r}
\end{align*}
and for $\deltab^{(i)}_h=1$
\begin{align*}
    &\underline{w}^{(i)}_{h,r}\leq \tilde{\xib}^{(i)}_{h,r} \leq \overline{w}^{(i)}_{h,r} \ \text{and} \ \tilde{\xib}^{(i)}_{h,r}= \tilde{\Wb}^{(i)}_{h,r},\\
    \forall h \in \{&1,\dots,p_i w_i\}, \ \forall r \in \{1,\dots,n\}, \  \forall i \in \{1,\dots,\ell\}.
\end{align*}
We can now rewrite the equality constraint in the third line of \eqref{eq:MI_Constraints_Gain} as follows
\begin{align*}
    \xib^{(i)}_{1,r}&=\sum\limits_{h=1}^{p_i} \deltab^{(i)}_h \tilde{\Wb}^{(i)}_{h,r}=\Delta^{(i)}_{1,:} \tilde{\Wb}^{(i)}_{:,r},\\
    \hspace{1cm}\vdots\\
    \xib^{(i)}_{w_i,r}&=\sum\limits_{h=p_i(w_i-1)+1}^{p_i w_i} \deltab^{(i)}_h \tilde{\Wb}^{(i)}_{h,r}=\Delta^{(i)}_{w_i,:} \tilde{\Wb}^{(i)}_{:,r}, \\ 
    \forall r &\in \{1,\dots,n\}, \ \forall i \in \{1,\dots,\ell\}.
\end{align*}
This can be written in a more compact form as matrix multiplication
\begin{equation*}
    \xib^{(i)}=\Delta^{(i)} \tilde{\Wb}^{(i)}=\Delta^{(i)} \Wb^{(i)} \xib^{(i-1)}, \ \forall i \in \{1,\dots,\ell\}.
\end{equation*}
Starting with $\xib^{(0)}=\Ib$, this recursion leads to
\begin{align*}
    \xib^{(\ell)}=\prod\limits_{i=1}^{\ell} \Delta^{(i)}\Wb^{(i)}.
\end{align*}
By substituting the former relation in \eqref{eq:K_NN_delta} we can show that the local gain is indeed given by \eqref{eq:K_NN_xi}, which completes the proof.
\end{proof}

The constraints \eqref{eq:MI_Constraints} and \eqref{eq:MI_Constraints_Gain} can now be used to model a maxout NN by MI linear constraints and to compute the output \eqref{eq:Maxout_q} and the local gain \eqref{eq:K_NN_xi} by solving an MI feasibility problem. If we further combine the results of Lemma~\ref{lem:MI_Maxout} and \ref{lem:MI_Maxout_gain}, we can compute the maximum error and the $\alpha$-Lipschitz constant of the error.

\begin{thm}\label{thm:errorMILP}
Let $\Phib(\xb)$ be a maxout NN and $\alpha=\{1,\infty\}$. Then the maximum error \eqref{eq:MaxError} and the $\alpha$-Lipschitz constant of the error \eqref{eq:Lipschitz_e} can be computed by solving an MILP.
\end{thm}
\begin{proof}
The proof follows from \citep[Thm.~6.1]{Fabiani2021}, where it is shown that both values can be computed by solving an MILP, if the output $\Phib(\xb)$ and the local gain $\Kb_{\text{NN}}(\xb)$ of the NN can be computed by solving an MI feasibility problem. This is always possible for maxout NN according to Lemma~\ref{lem:MI_Maxout} and \ref{lem:MI_Maxout_gain}. Thus, by replacing the MI linear constraints for the ReLU NN in \citep[Thm.~6.1]{Fabiani2021} with the constraints \eqref{eq:MI_Constraints} and \eqref{eq:MI_Constraints_Gain}, we can compute \eqref{eq:MaxError} and \eqref{eq:Lipschitz_e} for a maxout NN by solving an MILP.
\end{proof}

\section{Numerical examples}\label{sec:examples}

We consider two simple examples to highlight our theoretical findings.
In both examples, the maximum error \eqref{eq:MaxError} and the $\alpha$-Lipschitz constant of the error \eqref{eq:Lipschitz_e} are computed by solving the MILP according to Theorem~\ref{thm:errorMILP} with the MOSEK optimization toolbox for MATLAB (see \citep{mosek}). The NN are implemented in Python with Keras \citep{keras} and Tensorflow \citep{tensorflow}. During the training, the MSE \eqref{eq:MSE} is minimized with respect to the weighting matrices and bias vectors of the NN using a stochastic gradient descent.

\subsection{Example system with $n=1$}

We consider the system from \citep[Ex.~2]{SchulzeDarup2016_ECC_MPC}
with the dynamics
\begin{equation}
    \nonumber
    x(k+1)=\tfrac{6}{5}x(k)+u(k)
\end{equation}
and the constraints $\Xc = [-10,10]$ and $\Uc = [-1,1]$.  As in \citep{SchulzeDarup2016_ECC_MPC}, we choose $Q=19/5$, $R=1$, $P=5$, and $\Tc=[-1,1]$. Finally, we select $N=2$ for illustration purposes here.
Explicitly solving the OCP \eqref{eq:OCP} then leads to the control law
\begin{equation}\label{eq:example1_pi}
    \pi(x) = \left\{
    \begin{array}{lll} \vspace{1mm}
        -1 & \text{if} & x\in\left[-\tfrac{20}{9},-1\right], \\ \vspace{1mm}
        -x & \text{if} & x\in\left[-1,1\right], \\
        \blind{-}1 & \text{if} & x\in\left[1,\tfrac{20}{9}\right].
    \end{array}
    \right. 
\end{equation}
For a maxout NN with the topology from Corollary~\ref{cor:exactMaxout} (i) with $p_1=2$,
\begin{align}
\nonumber
    \Wb^{(1)}&=\begin{pmatrix}
        -1 & 0 & -1 & 0
    \end{pmatrix}^\top, \ 
    \bb^{(1)}=\begin{pmatrix}
        0 & -1 & -1 & 0
    \end{pmatrix}^\top \ \text{and} \\ 
    \label{eq:exact_W_1d}
    \Wb^{(2)}&=\begin{pmatrix}
        1 & -1
    \end{pmatrix} 
\end{align}
we have
\begin{equation}
    \label{eq:PhiEqualsPi1d}
    \Phi(x)=\max\{-x,-1\}-\max\{-x-1,0\}=\pi(x).
\end{equation}
We implemented the constraints \eqref{eq:MI_Constraints} and \eqref{eq:MI_Constraints_Gain} with $\overline{b}^{(i)}=\overline{w}^{(i)}_{h,r}=-\underline{w}^{(i)}_{h,r}=10^4$, $\varepsilon=10^{-5}$ and used them to compute \eqref{eq:MaxError} and \eqref{eq:Lipschitz_e} with $\alpha=\infty$ according to Theorem~\ref{thm:errorMILP}. For a maxout NN with the parameters \eqref{eq:exact_W_1d}, which exactly represents the control law, we computed $\overline{e}_\infty=4.5 \times 10^{-16}$ and $\Lc_\infty(e,\Tc)=0$. This can be seen as an experimental evidence that the implementation computes correctly the maximum error and the $\infty$-Lipschitz constant of the error. 

To obtain the data in Table~\ref{tab:1d}, we sampled randomly $1000$ points from the control law \eqref{eq:example1_pi} and trained different maxout NN for $1000$ epochs.     
\begin{table}[h]
    \caption{Training results of different maxout NN with $\ell=1$ and $w_2=1$ for $n=1$.}
    \centering
    \begin{tabular}{ccccccc}
        \toprule
        No.\!\!\!\! & $w_1$\!\!\!\! & $p_1$\!\!\!\! & $\#_p$\!\!\!\! & $\hat{e}$\!\!\!\! & $\overline{e}_\infty$\!\!\!\! & $\Lc_\infty(e,\Tc)$ \\  
        \midrule
        1.\!\!\!\! & $1$\!\!\!\! & $4$\!\!\!\! & $10$\!\!\!\! & $0.19$\!\!\!\! & $0.58$\!\!\!\! & $0.74$ \\[.5ex]
        2.\!\!\!\! & $2$\!\!\!\! & $4$\!\!\!\! & $19$\!\!\!\! & $1.39 \times 10^{-6}$\!\!\!\! & $3.22 \times 10^{-6}$\!\!\!\! & $2.24 \times 10^{-7}$ \\[.5ex]
        3.\!\!\!\! & $2$\!\!\!\! & $3$\!\!\!\! & $15$\!\!\!\! & $1.87 \times 10^{-6}$\!\!\!\! & $4.36 \times 10^{-6}$\!\!\!\! & $2.04 \times 10^{-7}$ \\[.5ex]
        4.\!\!\!\! & $2$\!\!\!\! & $2$\!\!\!\! & $11$\!\!\!\! & $1.48 \times 10^{-6}$\!\!\!\! & $3.26 \times 10^{-6}$\!\!\!\! & $1.69 \times 10^{-6}$ \\[.5ex]
        5.\!\!\!\! & $3$\!\!\!\! & $2$\!\!\!\! & $16$\!\!\!\! & $1.33 \times 10^{-6}$\!\!\!\! & $3.73 \times 10^{-6}$\!\!\!\! & $1.73 \times 10^{-6}$ \\[.5ex] 
        6.\!\!\!\! & $4$\!\!\!\! & $2$\!\!\!\! & $21$\!\!\!\! & $1.26 \times 10^{-6}$\!\!\!\! & $2.93 \times 10^{-6}$\!\!\!\! & $3.01 \times 10^{-7}$ \\[.5ex]
        7.\!\!\!\! & $4$\!\!\!\! & $1$\!\!\!\! & $13$\!\!\!\! & $0.24$\!\!\!\! & $0.45$\!\!\!\! & $0.65$ \\
        \bottomrule
        \label{tab:1d}
    \end{tabular}
\end{table}

The results show that a maxout NN with $w_1=2$ and $p_1=2$, which can represent the control law exactly (cf. \eqref{eq:PhiEqualsPi1d}), is sufficient to get a good approximation, in terms of maximum error and $\infty$-Lipschitz constant of the error. We only get an approximation with a relatively high error for the first and last topology of Table~\ref{tab:1d}. This is not surprising since for $w_1=1$, $p_1=4$ the maxout NN is a convex PWA function and for $w_1=4$, $p_1=1$ it is an affine function. In both cases, it is not possible to find an approximation of \eqref{eq:example1_pi} with a maximum error close to zero. 

\subsection{Example system with $n=2$}

We consider the OCP \eqref{eq:OCP} with
\begin{align*}
    \label{eq:ex2D}
    \Ab&=\begin{pmatrix}1 & 1 \\0 & 1\end{pmatrix}, \quad \Bb = \begin{pmatrix}0.5 \\1\end{pmatrix}, \\
    \Xc &= \{\xb\in\R^2 \ | \ |\xb_1| \leq 25, \ |\xb_2| \leq 5 \}, \\
    \Uc &= \{u\in\R \ | \ |u| \leq 1\},
\end{align*}
and choose $\Qb=\Ib$, $R=1$, $N=3$, $\Pb$ as the solution of the discrete-time algebraic Riccati equation and $\Tc$ as the maximal output admissible set (see \citep{Gilbert1991} for details). Explicitly solving the OCP leads to a control law with $29$ regions. Using the procedure described in \citep[Sec.~1]{Kripfganz1987}, we find a maxout NN of the type (i) from Corollary~\ref{cor:exactMaxout} with $p_1=38$ and $231$ parameters, that exactly describes the control law. For this NN we computed $\overline{e}_\infty=2.68 \times 10^{-12}$ and $\Lc_\infty(e,\Tc)=3.11 \times 10^{-5}$. A ReLU NN that can exactly represent the control law has according to \citep[Thm.~1]{Karg2020} $457$ parameters. This shows that, as already stated in Section~\ref{sec:MaxoutNNaMPC}, ReLU NN for the exact representation of PWA functions are more conservative compared to maxout NN. 

Table~\ref{tab:2d} summarizes the training results for different maxout NN trained for $1000$ epochs on $10^4$ samples of the control law. The values of \eqref{eq:MaxError} and \eqref{eq:Lipschitz_e} are computed as in the first example, except that in this example $\varepsilon=10^{-3}$ is chosen.  
\begin{table}[h]
    \caption{Training results of different maxout NN with $\ell=1$ and $w_2=1$ for $n=2$.}
    \centering
    \begin{tabular}{ccccccc}
        \toprule
        No.\!\!\!\! & $w_1$\!\!\!\! & $p_1$\!\!\!\! & $\#_p$\!\!\!\! & $\hat{e}$\!\!\!\! & $\overline{e}_\infty$\!\!\!\! & $\Lc_\infty(e,\Tc)$ \\ 
        \midrule
        1.\!\!\!\! & $2$\!\!\!\! & $38$\!\!\!\! & $231$\!\!\!\! & $4.47 \times 10^{-3}$\!\!\!\! & $4.59 \times 10^{-2}$\!\!\!\! & $2.33 \times 10^{-3}$ \\[.5ex]
        2.\!\!\!\! & $2$\!\!\!\! & $10$\!\!\!\! & $63$\!\!\!\! & $6.04 \times 10^{-3}$\!\!\!\! & $1.26 \times 10^{-1}$\!\!\!\! & $6.24 \times 10^{-3}$ \\[.5ex]
        3.\!\!\!\! & $2$\!\!\!\! & $3$\!\!\!\! & $21$\!\!\!\! & $2.45 \times 10^{-2}$\!\!\!\! & $1.69 \times 10^{-1}$\!\!\!\! & $1.31 \times 10^{-2}$ \\[.5ex]
        4.\!\!\!\! & $3$\!\!\!\! & $3$\!\!\!\! & $31$\!\!\!\! & $2.19 \times 10^{-2}$\!\!\!\! & $1.62 \times 10^{-1}$\!\!\!\! & $7.78 \times 10^{-2}$ \\[.5ex]
        5.\!\!\!\! & $5$\!\!\!\! & $3$\!\!\!\! & $51$\!\!\!\! & $1.97 \times 10^{-2}$\!\!\!\! & $1.63 \times 10^{-1}$\!\!\!\! & $9.60 \times 10^{-2}$ \\[.5ex]
        6.\!\!\!\! & $10$\!\!\!\! & $3$\!\!\!\! & $101$\!\!\!\! & $1.31\times 10^{-2}$\!\!\!\! & $1.30 \times 10^{-1}$\!\!\!\! & $2.05 \times 10^{-1}$ \\[.5ex]
        7.\!\!\!\! & $23$\!\!\!\! & $3$\!\!\!\! & $231$\!\!\!\! & $8.95 \times 10^{-3}$\!\!\!\! & $9.94 \times 10^{-2}$\!\!\!\! & $4.27 \times 10^{-1}$ \\
        \bottomrule
        \label{tab:2d}
    \end{tabular}
\end{table}

The experimental data indicate that the first maxout NN is the best choice. Note that this is a maxout NN with the topology (i), which theoretically allows an exact representation of the control law, trained on samples of the control law. The seventh maxout NN is of the type (ii), where $w_1$ is chosen such that the number of parameters is equal to that of the first maxout NN. It has a comparable maximum error but a $\infty$-Lipschitz constant that is about $183$-times higher. This observation may be explained by the fact that maxout NN with more neurons represent PWA functions with more regions compared to maxout NN with less neurons (see \citep[Thm.~3.6]{Montufar2021}). This makes it more likely that there exits one region for which the deviation between the local gains of the NN and the control law is larger, resulting in a larger $\infty$-Lipschitz constant (cf. \eqref{eq:Lipschitz_e}). The increasing $\infty$-Lipschitz constant with the number of neurons $w_1$ in Table~\ref{tab:2d} supports this assumption.            

\section{Conclusions and outlook}\label{sec:conclusions}

We presented a method to compute two values, i.e., the maximum error \eqref{eq:MaxError} with respect to the optimal control law and the Lipschitz constant \eqref{eq:Lipschitz_e} of the error function \eqref{eq:error}, which are crucial to certify stability of the closed-loop system with a maxout NN controller that approximates the optimal control law of MPC. Our results are derived by showing how the output and the local gain of maxout NN can be computed by solving an MI feasibility problem (cf. Lems.~\ref{lem:MI_Maxout} and \ref{lem:MI_Maxout_gain}). The combination of both lemmas leads to Theorem~\ref{thm:errorMILP}, which states that the maximum error and the $\alpha$-Lipschitz constant of the error function can be computed by solving an MILP. The computation of both values has been successfully applied to a number of realizations, 
including maxout NN which exactly describe the control law. 

An interesting direction for future research is the combination of the proposed method with the results from \citep{Teichrib2022}, where maxout NN are designed that allow an exact description of the piecewise quadratic optimal value function in MPC. Such a combination may provide a new method to analyze maxout NN approximations of the optimal value function.


\end{document}